\documentclass[12pt]{iopart}
\usepackage{graphicx}
\usepackage{dcolumn}
\usepackage{bm}

\usepackage{iopams}
\def\be{\begin{equation}}
\def\ee{\end{equation}}
\def\ba{\begin{array}}
\def\ea{\end{array}}
\def\bea{\begin{eqnarray}}
\def\eea{\end{eqnarray}}
\begin{document}
\title[\underline{J. Phys. G: Nucl. Part. Phys. \hspace {6cm} S. K. Patra {\it et al}}]
{Anatomy of neck configuration in fission decay}
\author{S.K. Patra$^{1}$, R. K. Choudhury$^{2}$ and L. Satpathy$^{1}$} 
\address{$^1$Institute of Physics, Sachivalaya Marg, Bhubaneswar 751005, India\\
$^{4}$Bhaba Atomic Research Centre, Mumbai}

\begin{abstract}

The anatomy of neck configuration in the fission decay of Uranium and Thorium
isotopes is investigated in a microscopic study using Relativistic mean field
theory. The study includes $^{236}U$ and $^{232}Th$ in the valley of stability
and exotic neutron rich isotopes $^{250}U$, $^{256}U$, $^{260}U$, $^{240}Th$,
$^{250}Th$, $^{256}Th$ likely to  play important role in the r-process
nucleosynthesis in stellar evolution. Following the static fission path, the neck
configurations are generated and their composition in terms  of the number of
neutrons and protons are obtained showing the progressive rise  in the neutron
component with the increase of mass number. Strong correlation between the neutron
multiplicity in the fission decay and the number of neutrons in the neck is 
seen. The maximum neutron-proton ratio is about 5 for $^{260}$U and $^{256}$Th
suggestive of the break down of liquid-drop picture and inhibition of the fission 
decay in still heavier isotopes. 
Neck as precursor of a new mode of fission decay like multi-fragmentation 
fission
may also be inferred from this study.
\end{abstract}



\section{Introduction}

Although the phenomenon of nuclear fission has been discovered since about sixty years
back, many crucial facets of this process are still not understood at 
fundamental level,
warranting further exploration of its dynamics. This is more so when viewed  from the
microscopic theoretical point of view. From the early days,   the two goals of nuclear
theory were to explain the fission phenomena and the working of the nuclear 
independent-particle shell model
in terms of the nuclear Hamiltonian involving  nucleon-nucleon interaction. Although the
later goal is more or less reached, the former is still languishing in the mid way.
The general methods followed in such studies are nonrelativistic selfconsistent
mean-field theories like Hartree-Fock (HF), Hartree-Fock Bogolybuv(HFB), Hartree-BCS
etc. and their derivatives.  These  theories are more suitable for finding 
the static fission
path \cite{x1,nx1,nx2,x2}, however they can
be used to exhaustively map the entire potential energy landscape spanned
by relevant degrees of freedom revealing main features of fission dynamics.
In an  improved microscopic calculation fission is described by performing the
time evolution of a single determinantal many-body wave function  in the framework of time-dependent 
Hartree-Fock (TDHF) theory \cite{x3}. Such attempts imposing restrictive constraints,
have been done with limited success. The proper microscopic theory for fission is the
adiabatic time-dependent Hartree-Fock (ATDHF) developed by Villars and others 
\cite{x4,x5} to deal with
effectively the large amplitude slow evolution of the nuclear shape in this process. In
this theory, one can calculate the dynamic fission path without computing the entire energy
landscape. However the practical application to date has been rather very 
limited  \cite{x5} primarily
because of its complexity. In a recent development a new microscope approach based on Time Dependent
Generator Coordinate Method (TDGCM) has been proposed \cite{nx3} using solutions of
Hartree-Fock Bogolyubov mean field theory and Gaussian overlap approximation.
The theory is equipped to describe the time-dependent evolution of the fission
process whose initial applications look promising.
The most  popular and widely used 
microscopic-macroscopic method \cite{nx4} is semi-microscopic in nature, consisting of a liquid drop part and
a shell correction part for the total energy of the nucleus. The exploration of the
multidimensional energy landscape has been quite successful in this method
\cite{x6}. In such approach, specifically introducing neck degree of freedom and using 
two-centre microscopic potential to favour the formation of two fragments, 
studies \cite{nx5,nx6} have been made to investigate specific questions like how neck influences the
fission paths and emission of scission neutrons etc.   Here we have
entertained a specific objective of studying the neck configuration following the static
fission path at a fundamental level using microscopic nuclear many-body Hamiltinian, with 
the hope of unraveling some core features of fission dynamics, for which
we have adopted the relativistic mean field (RMF) theory.

The critical feature of the fission phenomena is the multiplicity of neutron 
emitted in this
process, which plays key role in the chain reaction leading to energy generation. In the
commonly used thermal fission of $^{235}U$, the multiplicity of the neutron is 2.5, emitted
by the two fragments in the post scission stage after they are accelerated 
by the mutual
Coulomb repulsion. In this reaction, the neck is believed to be neutron rich as the neutron
emission from this region have been observed \cite{x7,x8} to be more than 
that of the $\alpha$- particle
and the proton emission. Experimentally it may not be possible to ascertain the true
composition of the neck, which has potential to reveal important aspects of the mechanism
of fission dynamics. It will be truly rewarding if it would be possible to generate
theoretically the neck, and find neutron-proton composition in a quantitative manner. Such
study will open up the possibility for  understanding the fission of highly neutron-rich
actinide nuclei, likely to be synthesised in the RIB fascilities under construction in many
laboratories around the world \cite{tho67}. How does such a nucleus undergo fission is a very pertinent
question with potential to be an exciting theme of research in future. The  composition of
neck in the fission of such nuclei may contain informations  regarding new physics. Various
mass formulae have predicted that Uranium isotope with mass number as high 
as 270 and even more
upto 290 to be particle stable. The progressive production of such nuclei in the r-process
of nuclear synthesis in stellar evolution is more real than speculative. The continued
synthesis of successively heavier isotopes of an element by neutron capture in this process
is finally terminated by fission. In an earlier study \cite{x9}, we have 
identified a valley where
Uranium and Thorium isotopes with neutron number in  the range N=154-172 to be thermally
fissile, a reflection  of the close shell structure of N= 162 $\sim 164$ predicted in numerous
calculations over the years. Therefore in our calculation we have included, besides the two
well known stable nuclei $^{232}$Th and $^{236}$U the exotic nuclei $^{250}U$, $^{256}U$, $^{260}$U,
$^{240}Th$, $^{250}Th$ and $^{256}$Th  far from this stability valley.

A more exciting impetus to study the above nuclei originates from the expectation to find
new phenomena in this virgin nuclear territory of high neutron to proton ratio,
 widely speculated to
be manifested. The maximum neutron to proton ratio of neck found to be about
5 for the heaviest isotopes $^{260}$U and $^{256}$Th considered here.
The heavier isotopes than these ones will not support neck suggestive of 
the break down of the liquid drop picture. Our recent study \cite{x9} 
of $^{250}$U gives strong evidence of new mode of fission decay termed
as 'multifragmentation fission' in which 2.5 neutrons are emitted at scission along with the two
fragments, in addition to the usual 2.5 prompt  neutrons in the post scission stage. The present
study of the composition of the neck configuration has the potential to be a precursor of such a
phenomenon and even of some more exotic ones. In Sec.2 , we present an outline of our scheme of
calculation in the RMF theory.The calculations and results are given in Sec.3, and discussions in
Sec.4. Sec. 5 contains the conclusion.

\section{Scheme of calculation in RMF theory}
A brief sketch of the RMF theory \cite{4,sero86} is outlined
here for completeness.  
The relativistic Lagrangian density for a nucleon-meson many-body 
system is given by

\begin{eqnarray}
{\cal L}&=&\overline{\psi_{i}}\{i\gamma^{\mu}                                   \partial_{\mu}-M\}\psi_{i}
+{\frac12}\partial^{\mu}\sigma\partial_{\mu}\sigma                              -{\frac12}m_{\sigma}^{2}\sigma^{2}\nonumber\\
&& -{\frac13}g_{2}\sigma^{3} -{\frac14}g_{3}\sigma^{4}                          -g_{s}\overline{\psi_{i}}\psi_{i}\sigma-{\frac14}\Omega^{\mu\nu}
\Omega_{\mu\nu}\nonumber\\                                                      &&+{\frac12}m_{w}^{2}V^{\mu}V_{\mu}
+{\frac14}c_{3}(V_{\mu}V^{\mu})^{2} -g_{w}\overline\psi_{i}                     \gamma^{\mu}\psi_{i}
V_{\mu}\nonumber\\                                                              &&-{\frac14}\vec{B}^{\mu\nu}.\vec{B}_{\mu\nu}+{\frac12}m_{\rho}^{2}{\vec
R^{\mu}} .{\vec{R}_{\mu}}
-g_{\rho}\overline\psi_{i}\gamma^{\mu}\vec{\tau}\psi_{i}.\vec
{R^{\mu}}\nonumber\\
&&-{\frac14}F^{\mu\nu}F_{\mu\nu}-e\overline\psi_{i}
\gamma^{\mu}\frac{\left(1-\tau_{3i}\right)}{2}\psi_{i}A_{\mu}.
\end{eqnarray}

From this Lagrangian we obtain the field equations for the nucleons and mesons.
These equations are solved by expanding the upper and lower components of 
the Dirac spinors and the boson fields in an axially deformed harmonic 
oscillator basis.  The set of coupled equations is solved numerically by 
a self-consistent iteration method. Here we get bosonic equation for $\sigma$, 
$\omega$ and $\rho$ mesons and Dirac equation for the nucleons.  
The $\sigma$- field gives the attractive and
the $\omega$- field gives the repulsive component of the nuclear potential. 
However,
the $\rho$  meson field takes care of the proton-neutron asymmetric energy. In our
calculation, the constant gap BCS pairing is used to add the pairing effects
for open shell nuclei. The
centre-of-mass motion (c.m.) energy correction is estimated by the usual harmonic
oscillator formula $E_{c.m.}=\frac{3}{4}(41A^{-1/3})$. 

In the present investigation, we have carried out study
microscopically in the nonlinear RMF theory of Boguta and Bodmer \cite{12},
an extended version of Walecka \cite{4} theory. We have adopted the 
NL3 \cite{lala97} interactions. The NL3 interaction has been widely 
used in recent
years in the calculation of varieties of nuclear properties like
binding energy, {\it rms} radii and giant resonances etc. and have been
accepted to be very successful. Here we have studied the
stability of our result for each nucleus by varying the number of 
harmonic oscillator shells used in the basis between $N_F=N_B=12$ to 28.
The quadrupole deformation parameter $\beta_2$ is evaluated from the resulting 
proton and neutron quadrupole moments, as $Q=Q_n+Q_p=\sqrt{\frac{16\pi}5} 
(\frac3{4\pi} AR^2\beta_2)$. The {\it rms} matter radius is 
defined as $\langle r_m^2\rangle={1\over{A}}\int\rho(r_{\perp},z) r^2d\tau$; 
here $A$ is the mass number, and $\rho(r_{\perp},z)$ is the deformed density. 
As outputs, we obtain different potentials, 
densities, single-particle energy levels, radii, quadrupole deformations 
and the binding energies. For a given nucleus, the maximum binding energy corresponds to
the ground-state and other solutions are obtained as various excited intrinsic states, 
provided the nucleus does not under go fission.  The density distribution of nucleons 
plays the prominent role in the study of the internal structure of a nucleus. 
When the deformation of a nucleus is varied, the density distribution 
$\rho(r_{\perp}, z)$ inside 
the nucleus also varies. For example, the $\rho(r_{\perp}, z)$ for a spherical nucleus is 
symmetrical in $(r_{\perp}, z)-$plane. However, it is highly asymmetric for
a largely deformed nucleus. Knowing the density distribution of the spherical or (oblate/ prolate) 
deformed configuration, we can calculate the number of nucleons for each 
configuration, defined as 
\begin{equation}
N=\int\int\rho_n(r_{\perp}, z) d\tau,  
\end{equation}
\begin{equation}
Z=\int\int\rho_p(r_{\perp}, z) d\tau,  
\end{equation}

\noindent where $\rho_n$ and $\rho_p$ are the neutron and proton density 
of the nucleus, and the the total density $\rho=\rho_n+\rho_p$.
We have obtained the static fission path by calculating the potential 
energy surface (PES) using the above RMF formalism in a 
constrained calculation \cite{patra09,flocard73,koepf88,reinhard89}, 
i.e., instead of minimizing the $H_0$, we have minimized 
$H'=H_0-\lambda Q_{2}$, with $\lambda$ 
as a Lagrange multiplier and $Q_2$, the quadrupole moment. 
Thus, we get the solution at a given quadrupole deformation and
then obtain the energy using the $H_0$.

\section{Calculation and Result}

The fission process has been traditionally visualized in liquid-drop picture,
where the nuclear surface is expressed in terms of collective deformation
parameters characterized by several multipole shapes with quadrupole deformation
playing the most important role. Classically, it corresponds to slow movement of the
nucleus from the ground state equilibrium shape along a slowly rising valley
path in the energy landscape spanned by the deformation variables. It reaches
the barrier top (saddle point),  and then slides down on the outer surface of the
barrier and acquires a highly elongated shape culminating in a narrow neck which
finally splits into two fragment nuclei. Quantum mechanically
this path upto the scission point defines a potential barrier, which can be
used for calculating the transmission coefficient to obtain the fission
probability. It is this path normally called static fission path, which
we intend to calculate here. Since quadrupole deformation plays the predominant
role, we have chosen it in our calculation and ignored the other deformation coordinates
for simplicity, being  guided by the goal for studying the anatomy of neck configurations
exhaustively.

It is worth pointing out that a nucleus undergoes binary fission in various
modes with different probabilities. The neck configuration in each mode is
likely to be different from those of the other modes depending upon the respective
pairs of fragment nuclei. Therefore, the necks for the symmetric and asymmetric
configurations will be different, and so also their scission points.

In our calculations we have imposed time reversal symmetry, reflection symmetry across XY plane
and rotational symmetry about the Z-direction, i,e, the fission
direction. So naturally we arrive at symmetric fission.
M\"oller et al.\cite{x6} have found two different paths leading to symmetric
and asymmetric fission in their study in five-dimensional deformation space.
We did not get solution for the
configuration with two separated fragments in our calculations, which may be attributed
to the limitation of the numerical computation/mean field theory.

In Table 1, we have presented our results on the energy,
deformation, rms radius of the ground-state and neck configuration 
of the two known nuclei $^{236}$U and $^{232}$Th in the valley of stability
and the six  exotic neutron-rich nuclei $^{250}$U, $^{256}$U, $^{260}$U,
$^{240}$Th, $^{250}$Th and $^{256}$Th away from it. To begin with it 
will be interesting
and reassuring to see how well the ground-state properties of these two
known nuclei are reproduced in our calculation. The experimental values of
these properties are presented in parenthesis below the corresponding
theoretical values in the same table. It can be seen that the ground-state
energy, the deformation and rms radius of these two nuclei compare remarkably
well with experiment. This reassures us about the goodness of the interaction and the 
suitability of the RMF theory for this study. 

In the next step, we calculate the static fission-path and examine if the
general fission properties like the barrier and its double-humped 
characteristics are reproduced. Although we have performed startic path 
calculation for all
the eight nuclei we present here only the results of the widely studied 
nucleus $^{236}$U, and one of the exotic nuclei, i.e., $^{250}$U. In Fig. 1
we have plotted the energy as a function of the deformation parameter 
for these two nuclei. From the figure, it can be seen that the fission barrier
for $^{236}$U comes out to be 6.95 MeV comparable to the experimental
value of 5.75 MeV \cite{moller95}. The fission barrier of 4.05 MeV in 
case of $^{250}$U obtained
in our calculation, is expectedly lower than that of $^{236}$U which agrees with
the Howard-M\"oller value of 4.3 MeV. The double-humped fission barriers 
in both cases have been reproduced. These results are indeed very satisfying.
We have followed the density profile of the evolving fragments along with the neck,
to findout a criterion for the delineation of the latter. We observe in general that, 
this profile has the
form of a Fermi distribution in the fragment region as expected, with the tail part
attaining a constant value. We have chosen the criterion that when the density falls
to $15 \%$ of the central density and remains constant along the
Z-coordinate, then neck is assumed to originate which merges on the other fragment
with the endpoint being defined with similar criterion.

We have presented in Fig. 2, the matter density distributions of
some typical configurations these two nuclei acquire, in their static  
fission-path
right upto the neck configuration. It is satisfying to see well defined dumbbell
shape   of neck reproduced in RMF study as solution of microscopic nuclear
many-body Hamiltonian, in agreement with age-old  picture of fission process
envisioned in classical  liquid drop model. The physical characteristics of the
necks of these systems emerging   from this study will be discussed along
which the other six nuclei aposteriori. In a recent microscopic 
study using constrained HFB method with Gogny interaction Dubray et al \cite{nx1},
obtained very elongated shape without clear-cut neck before the scission in Th and Fm 
isotopes, and found that these
fissioning systems energetically favoured splitting into two separate fragments
rather than develop an elongated shape with a neck. Banneau \cite{nx7} obtained similar results
for Fm isotope in his study using Skyrme-Hartree-Fock-BCS model. 

\begin{table*}
\caption{The RMF(NL3) results for binding energy BE, the quadrupole 
deformation parameter $\beta_{2}$ and {\it rms} radius for both the ground and
neck configurations 
for $^{236}U$, $^{250}U$, $^{256}U$, $^{260}U$, $^{232}Th$, $^{240}Th$, 
$^{250}Th$ and $^{256}Th$. The experimental values wherever known 
are given in parenthesis.
}
\begin{tabular}{|c|c|c|c|c|}
\hline
Nucleus & State & $\beta_{2}$ & $r_{c} (fm) $ & BE (MeV)\\
\hline
$^{236}U$ & Ground & 0.26 (0.28\cite{raman01})& 5.86 (5.84 \cite{angeli04}) & 1790.5 (1790.4 \cite{audi03})\\
 & Neck & 6.04 & 12.14 & 1808.5 \\
$^{250}U$ & Ground & 0.26 & 5.82 & 1856.5\\
 & Neck  & 5.52 & 11.91 & 1890.7\\
$^{256}U$ & Ground & 0.18 & 5.97 & 1880.6\\
 & Neck & 5.41 & 11.78 & 1914.7\\
$^{260}U$ & Ground & 0.15 & 5.99 & 1895.3\\
 & Neck & 5.37 & 11.74 & 1923.9\\
$^{232}Th$ & Ground & 0.26 (0.26\cite{raman01})& 5.83 (5.71 \cite{angeli04}) & 1767.1 (1766.7\cite{audi03})\\
 & Neck & 5.79 & 11.88 & 1789.0\\
$^{240}Th$ & Ground & 0.26 & 5.90 & 1806.3\\
 & Neck & 5.52 & 11.66 & 1825.2\\
$^{250}Th$ & Ground & 0.22 & 5.93 & 1846.3\\
 & Neck & 5.45 & 11.84 & 1872.2\\
$^{256}Th$ & Ground & 0.16 & 5.96 & 1868.8\\
 & Neck & 5.31 & 11.72 & 1891.5\\
\hline
\end{tabular}
\label{Table 1}
\end{table*}

Since our objective has been to critically study the neck configurations,
we have presented their matter
density distributions in Figures 3 and 4, obtained in 
our calculation of the four Uranium isotopes
and four Thorium isotopes respectively.
As noted above, the energies, rms radii and deformations of the neck
configuration and the ground-states for 
all the eight nuclei can be seen in Table 1. The five nuclei
$^{236}$U, $^{250}$U, $^{232}$Th, $^{240}$Th and $^{250}$Th have
similar ground-state deformation with $\beta_2$ values around 0.26. The
remaining three nuclei $^{256}$U, $^{260}$U and $^{256}$Th have
significantly lower $\beta_2$ values around 0.16 which may be a reflection
of the proximity of their neutron numbers to shell closure N=164. The
ground-state rms charge radii of all the nuclei are around 6 fm;
the corresponding radii of the neck configurations are nearly twice
being around 12 fm which are very much expected. As can be seen in the 
Table I, the neck configurations lie around 20 MeV
below the respective ground-states in conformity with expectation and in agreement with our
general notion of fission dynamics. From the matter density 
distributions in Figs. 3 and 4, it is clear that all the cases are of
symmetric fission.  To see how far our neck 
configuration in $^{236}$U conforms to reality,  we have calculated
the kinetic energy that will be generated if it breaks into two fragments.
It comes out to be 139 MeV which may be compared with the most probable 
value of 155 MeV observed in low energy fission of $^{236}$U corresponding to 
asymmetric mass split.  

\section{Characteristics of neck}

\subsection{Density and N/Z composition}

Our calculation yields the matter density $\rho$, the neutron density $\rho_{n}$ and
the proton density $\rho_{p}$ at every point in the body of system \cite{aru05}. 
The total number
of neutrons $N_{neck}$ and the number of protons $Z_{neck}$ contained 
in the neck, are obtained by
integrating the corresponding densities over its physical dimension of the neck using
Eq. (1).
The results are presented in Table 2. We have also
presented the  mean densities
$\overline{\rho_{n}}=\int{\rho_{n}} d\tau/d\tau$,
$\overline{\rho_{p}}=\int{\rho_{p}} d\tau/d\tau$, of the necks in the same.
As expected the ${\bar\rho_p}$ for both the elements remain similar for
all their isotopes being around 0.035 $fm^{-3}$. As for the ${\bar\rho_n}$
for both the systems the values generally increase with the increase of the
neutron number of the isotopes.  It can be seen that with the increase
of the neutron number in the isotope, the neutron to proton density ratio
${\bar\rho_{n}}/{\bar\rho_{p}}$ increases generally, as expected.  
For U$-$isotopes,
the ratio has increased from 1.70 for $^{236}U$ to 3.0 for $^{260}U$.
The corresponding number is 1.51 for $^{232}$Th to 2.73 for $^{256}$Th.

\begin{table*}
\caption{The Characteristics of neck configuration. The average
neutron and proton densities ${\bar \rho_n} (fm^{-3})$, and 
${\bar \rho_p (fm^{-3})}$ and
their ratios ${\bar \rho_n}/{\bar \rho_p}$, number of neutron and proton
$N_{neck}$ and $Z_{neck}$ contained in the neck, rms charge radius $r_c^{nk} (fm)$, length 
of the
neck $L_n (fm)$, the centre to centre distance $l_{cc} (fm)$ and 
tip to tip distance of the neck $L_t (fm)$ 
are presented. 
}
\begin{tabular}{|c|c|c|c|c|c|c|c|c|c|c|}
\hline
Nucleus &${\bar\rho_n}$& ${\bar\rho_p}$& ${\bar\rho_n/{\bar\rho_p}}$
& $Z_{neck}$ & $N_{neck}$ &$\frac{N_{neck}}{Z_{neck}}$& $r_{c}^{nk} $ 
& $L_{n} $ & $l_{cc} $ & $L_{t}$ \\
\hline
$^{236}U$ &0.058 &0.034&1.70& 0.95 & 2.42 &2.54& 12.14 & 6.28 & 21.88&37.48 \\
$^{250}U$ &0.074 &0.031&2.37& 0.86 & 3.23 &3.75& 11.91 & 5.82 & 21.56&37.30 \\
$^{256}U$ &0.085 &0.031&2.74& 0.82 & 3.77 &4.59& 11.78 & 5.46 & 21.30&37.14 \\
$^{260}U$ &0.093 &0.031&3.00& 1.03 & 5.11 &4.96& 11.74 & 4.29 & 21.18&37.07 \\
$^{232}Th$&0.062 &0.041&1.51& 0.66 & 1.71 &2.59& 11.88 & 5.18 & 21.80&38.42 \\
$^{240}Th$&0.071 &0.042&1.69& 0.69 & 2.47 &3.57& 11.86 & 4.76 & 21.44&38.12 \\
$^{250}Th$&0.081 &0.034&2.38& 0.62 & 2.94 &4.74& 11.82 & 4.68 & 21.10&37.52 \\
$^{256}Th$&0.090 &0.033&2.73& 0.68 & 3.78 &5.55& 11.72 & 4.52 & 20.74&36.96 \\
\hline
\end{tabular}
\label{Table 1}
\end{table*}

We have estimated the number of neutrons and protons contained in the neck
by using Eq. (1), which are presented in Table 2.
As we move from $^{236}U$ to $^{260}U$, the number of neck neutrons increases from
2.42 to 5.11 and the proton number almost does not change. Similar trend
is seen for Th isotopes with rise from 1.71 for $^{232}$Th to 3.78 for
$^{256}$Th. It may be observed that the value of $N_{neck}/Z_{neck}$ is some what different 
from that of ${\bar{\rho_n}}/{\bar{\rho_z}}$ in Table 2, because the effective volumes
for neutron and proton distributions are different, the latter being somewhat smaller than the former
due to its considerable low number. This high neutron-proton ratio is a precursor of a new mode of
fission decay which will be discussed aposteriori. 

The maximum ratio of
neutron to proton in the neck found by us is around 5 for the heaviest
isotopes $^{260}$U and $^{256}$Th considered by us. Quasi-bound/resonance
for $^{5}$H, $^{6}$H have been known; no such states for heavier isotope of
hydrogen above $^6$H have been observed. Neck can be considered a quasi-bound
transient state. Thus our finding of the maximum
N/Z ratio in the neck correlates with these transient states observed in
nature. The nuclear matter  inhabiting the neck region  in the heaviest isotope 
$^{260}$U considered here
has the neutron to proton ratio about 5:1 which corresponds to $^6$H system.
Since $^7$H system
is the likely limiting surviving resonance with half-life of $23 X 10^{-24}$ sec. only
and resonances of still
heavier hydrogen isotopes are unlikely to be found in nature, it is reasonable to conclude that
such nuclear liquid is not viable. Therefore neck formation in somewhat heavier than
$^{260}$U isotope requiring such nuclear liquid with higher neutron to proton ratio than 6:1 will
be unsustainable, which signals break-down of the usual liquid-drop picture of the fission
process visualized so far. Since neck can not be formed as the nuclear liquid-drop picture breaks
down and  so fission will be inhibited in such heavy neutron-rich isotopes.  
This effect is also manifested in progressive shrinking of the elongation of the neck with
rise of neutron number in heavier isotopes (see Table 2)
which may eventually disappear with increase of mass.
Thus the predominant mode of decay of such nuclei will be $\beta-$decay. The situation 
seems parallel to the case of superheavy element in the valley of stability where $\alpha-$decay
becomes more preferred mode of decay over fission.

\subsection{Size and Extension}
How far the nucleus extends in its neck configuration compared to its size in
its ground-state ? In Table 2, we have presented the charge radius
of the neck configuration $r_c^{nk}$, length of the neck $L_n$, i.e. 
the distance between the two facing surfaces
connected by  the neck, $l_{cc}$ the distance between the centre of the
two fragments and $L_{t}$  the distance between the two tips which
is a measure of the extension. The charge radii of the neck for all the
isotopes are around 12 fm (see Table 1). The length of the neck $L_n$
is around 5 fm for the two nuclei 
$^{236}$U and $^{232}$Th. The value decreases with the rise
of the neutron number reaching 4.29 fm for $^{260}$U and 4.52 fm for
$^{256}$Th.  Less elongated neck for heavier isotopes is a reflection of the
fact that, decrease of binding energy due to increase of neutron to proton 
ratio,
does not support a longer configuration but a shorter one.

The tip to tip distance $L_t$ is around 37 fm in all the cases. 
Thus, well defined  neck and fairly  extended mass distribution in this 
configuration
is evident in all cases. It is indeed interesting   to find,  very heavy
nuclei acquiring such extended dumbell configuration supported by
nucleon-nucleon force. It is useful to recall here that light
4n nuclei like $^{16}$O, $^{20}$Ne, $^{24}$Mg, $^{28}$Si etc.
exhibit very long physical states with linear $\alpha-$cluster
configuration\cite{brink66}.   

\subsection{Precursor of a new mode of fission decay}

Since neck is the most sensitive part where rupture takes
place giving rise to two heavy fission fragments, its composition will play a crucial
role in determining the fission dynamics. In widely studied $^{236}$U, the decay process is
binary; at the instant of the rupture of the neck, each of two protruding 
parts originating from the rupture of the neck is sucked in by the respective
 connecting 
fragments.  Then the two fragments move in opposite direction,
being repelled by the mutual Coulomb repulsion, and emit 2.5 neutrons in the process of
de-excitation to the respective ground-states. The sucking of the half of the neck by
a fragment is possible because of the presence of sizable number of protons
(as shown in the present calculations) in the neck providing attractive nuclear force.
However in the heavier isotopes like $^{250}$Th, $^{256}$Th, $^{250}$U, $^{256}$U and
$^{260}$U the proton components are relatively smaller compared to the 
neutron components.  The likely scenario in such a system is that
the triplet-triplet and singlet-singlet components of
nucleon-nucleon force which are repulsive in nature become dominant, and therefore sucking
in is unlikely to happen, and the ruptured neck may not be able to hold on all the
neutrons, and may simultaneously release them along with the production of the two
fragments.
This will be a new mode of fission decay as mentioned in the
introduction known as {\it multifragmentation fission}
proposed earlier by the authors \cite{x9}. In such cases, the multiplicity 
of  neutron will consist
of two parts: the usual prompt neutrons  plus the multifragmentation neutrons  generated
at the time of scission contributed by the neck. Taking into account the neck neutrons
calculated above, and assuming the modest value of 2.5 for prompt neutrons similar to that
of $^{236}$U, the multiplicity are expected to be  more than 4 for $^{250}$U, 
$^{256}$U, $^{260}$U, $^{250}$Th and $^{256}$Th. This prediction will get strong support from the
probability of fragment mass-yield as shown below.

The important driving force for the decay of a nucleus is the Q-value of the reaction.
The probability of fragment mass-yield in fission process is directly related to the Q-value
and temperature at the scission point. We therefore calculate the Q-value systematics
of the fission of the nucleus (A, Z) decaying to ($A_{1}, Z_{1}$) and ($A_{2}, Z_{2}$)
defined as:
\begin{equation}
{Q_{fiss.}(A, Z)} = BE ({A_{1}}, {Z_{1}}) + BE ({A_{2}}, {Z_{2}})-BE (A, Z)
\end{equation}

To avoid clumsiness we
have plotted the Q-values in Fig. 5 for the  6 isotopes 
$^{236,250,260}$U and $^{232,250,256}$Th
for the possible binary decays (A, Z)= ($A_{1}, Z_{1}$)+ ($A_{2}, Z_{2}$) 
taking even values of $Z_{1}$ in the range 39 to 46, with varying $A_{1}$, 
the complementary fragment ($A_{2}, Z_{2}$) being thereby fixed. Since the 
yield falls rapidly with the
decrease of the Q-value for an element, the values lying above 90\% of the heighest value
are only relevant.

For the present study we have used the masses predicted in the finite range 
droplet model (FRDM) \cite{moller95} and
infinite nuclear mass model (INM) \cite{nayak99}. We have chosen the latter mass model for its unique
success in describing the saturation properties \cite{nxx1} of INM, shell quenching \cite{nxx2}
in the large neutron N=82, 126 
shells and its long range predictive ability especially in the neutron rich side. For
each element the two vertical lines in the figures refer to the drip-lines predicted in
the two mass models which agree within $3 \sim 5$ neurons. Except
for the three isotopes $^{232}Th$, $^{236}U$ and $^{240}$Th in the valley of stability,
in the remaining five caces, the drip-lines fall within the Q-value distributions with a
few touching the outer fringes. To avoid clumsyness we depict in Fig. 5 the cases of 
6 isotopes only.  All the isotopes lying to the right of the drip-line
will be unstable against the instantaneous release of the neutrons from the fragments
at the scission. In the usual fission process of  $^{236}U$, neutrons are emitted from
the fragments after they are accelerated. But the five cases of heavy isotopes
 $^{250}U$, $^{256}U$, $^{260}U$,  $^{250}Th$ and $^{256}Th$ a certain number
of neutrons will be simultaneously emitted  along with the two heavy fragments.
It will
be interesting to see if the number of multifragmentation neutrons which can be estimated
taking the drip-line nucleus as the last surviving fragment, and compare 
the same with the
number of neutrons in the neck. It is expected that there will be a correlation
between the number of neck neutrons and the multifragmentation neutrons
which will be emitted by the fragments being populated beyond 
drip-line at scission. It is reasonable to suppose
the excess of neutrons beyond the drip-line are likely to lodged in the
neck. From the $Q-$value distributions of $^{260}$U it can be seen that about 5
multifragmentation neutrons will be emitted (ignoring the one or two units of
neutron difference between predictions of the two mass formulae) in 
most of the decay modes which agree with the neutron number in the neck
(see Table 2). 
In the case of $^{256}$Th the corresponding number is about 4 neutrons. Thus in the
fission process of exotic neutron-rich isotopes, the picture emerges that,
the neutrons beyond the drip-line contained in the fission fragments are
likely to take part in the neck formation.  It is interesting
to see that there is a qualitative agreement. It must be 
recognised
that these scission neutrons are the extra neutrons in addition to the normal multiplicity
of neutrons emitted from the fragments. Thus the total multiplicity will be more than doubled.
It will have serious implication in the chain reaction, the key process for energy generation
in fission. We would like to imphasize that although at the moment this new decay mode of
fission may not be realised in the laboratory, it is very likely occurring in
the r-process nucleosynthesis of terrestrial evolution.

\begin{figure}
\centering 
\includegraphics[width=0.60\textheight,clip=true,angle=0]{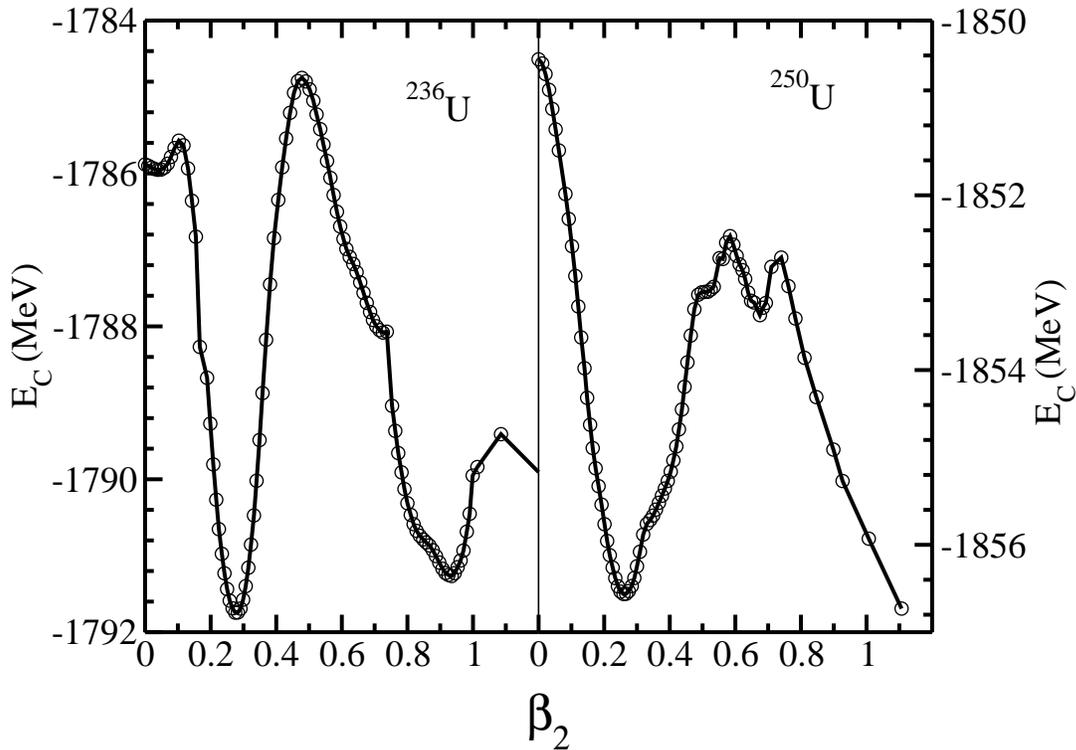}
\caption{The energy curves of $^{236}U$ and $^{250}U$ as a function of 
the deformation parameter $\beta_2$ obtained in the Relativistic
mean field formalism with NL3 parameter set.
}
\end{figure}

\begin{figure}
\centering 
\includegraphics[width=0.60\textheight,clip=true,angle=0]{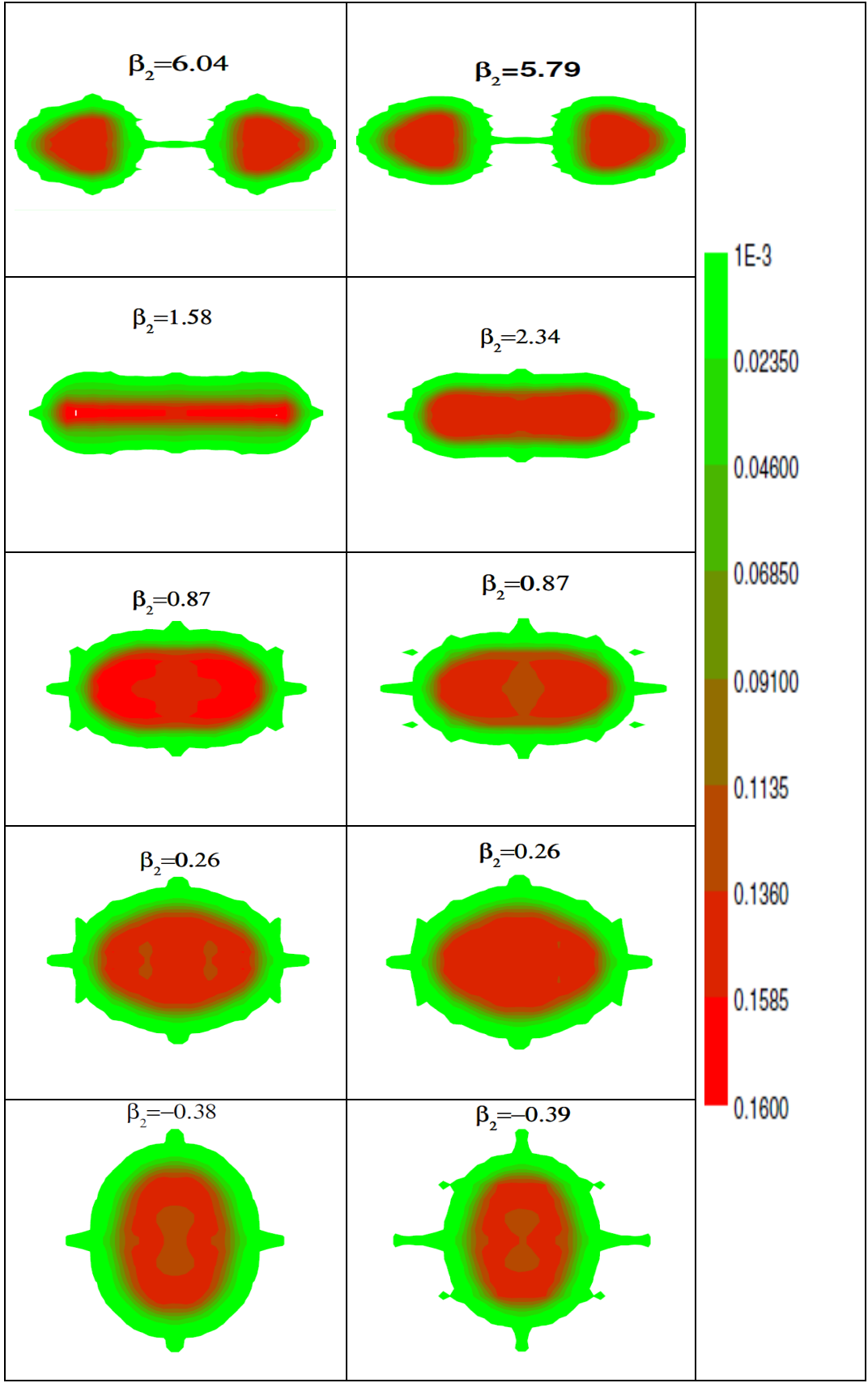}
\caption{Evolution of neck for the isotopes of
Uranium and Thorium. The matter density distributions for different
deformation $\beta_2$ of $^{236}$U and $^{232}Th$ obtained in
the relativistic mean field formalism using NL3 parameter set.
The total (neutron+proton) number density $\rho=\rho_n+\rho_p$ 
in $fm^{-3}$ is shown.}
\end{figure}

\begin{figure}
\centering 
\includegraphics[width=0.60\textheight,clip=true,angle=0]{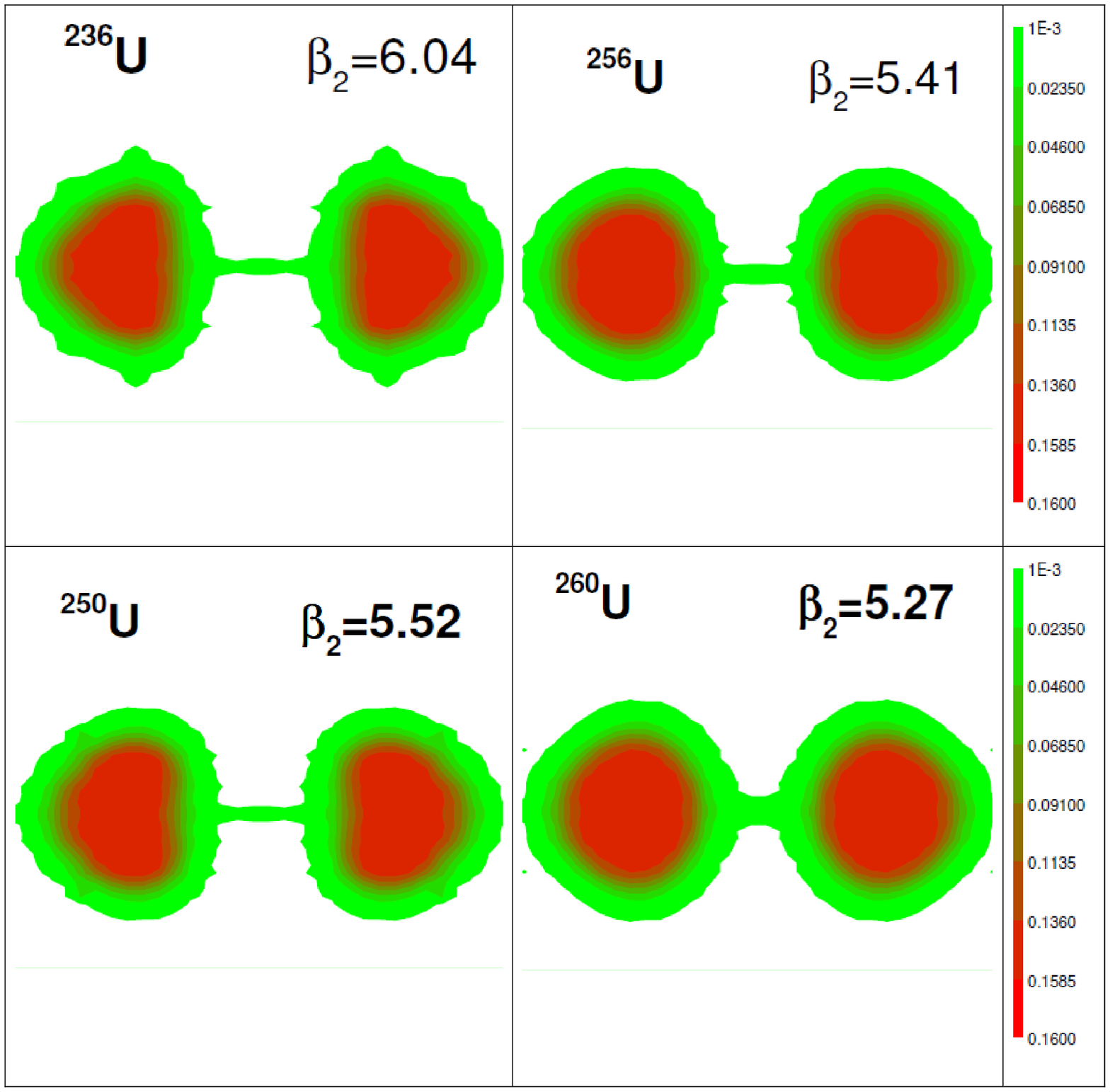}
\caption{The matter density distributions for the
neck structure of Uranium isotopes in a relativistic
mean field formalism using NL3 parameter set. The total (neutron+proton)
number density $\rho=\rho_n+\rho_p$ in $fm^{-3}$ is shown.}
\end{figure}

\begin{figure}
\centering 
\includegraphics[width=0.6\textheight,clip=true,angle=0]{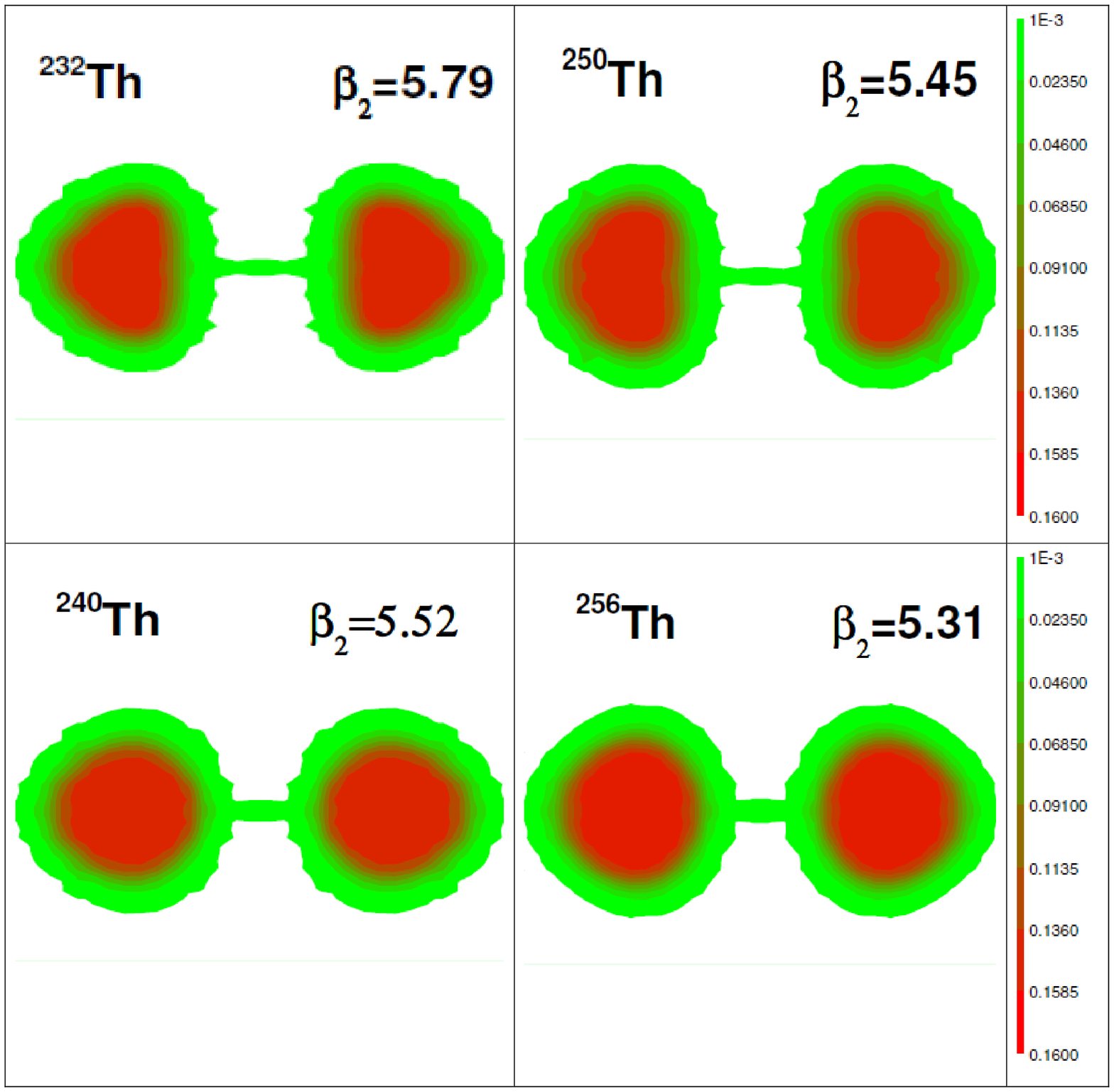}
\caption{Same as Figure 2, but for Thorium isotopes.
}
\end{figure}

\begin{figure}
\centering 
\includegraphics[width=0.65\textheight,clip=true,angle=0]{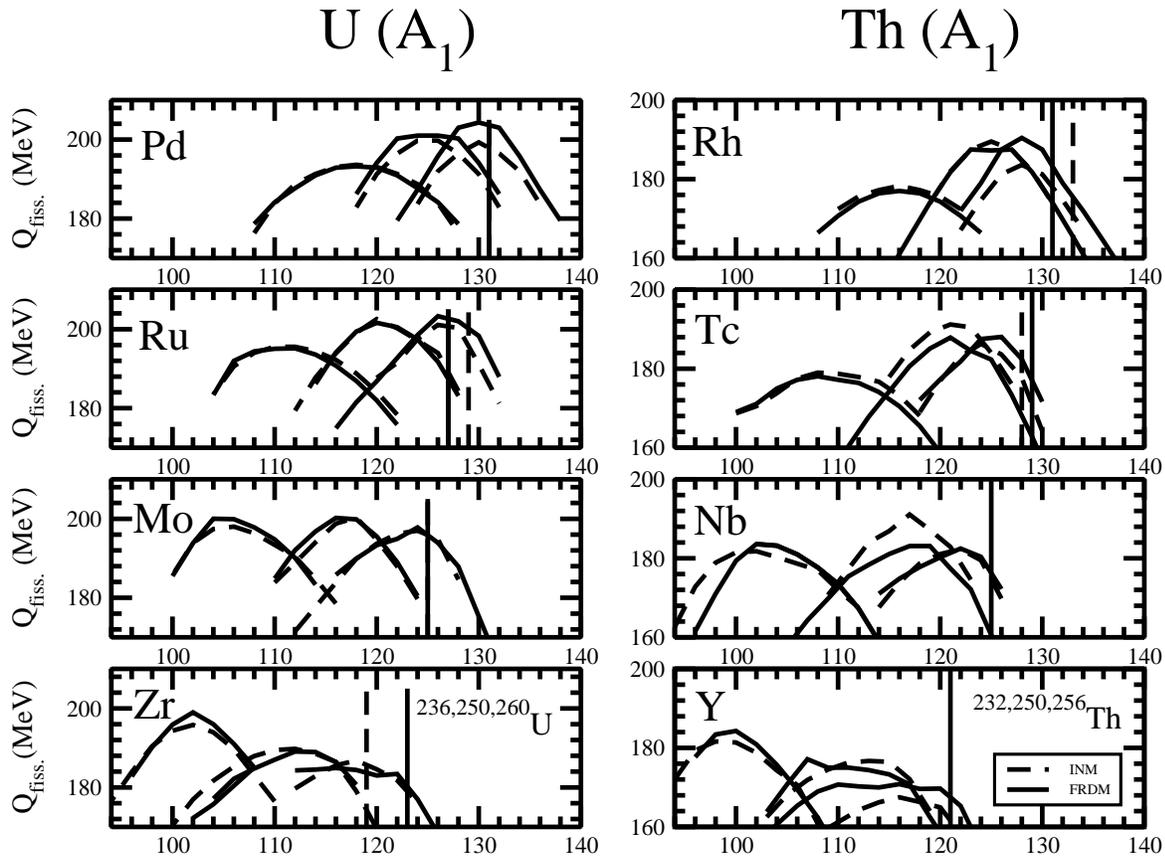}
\caption{ $Q_{fiss.}-$value distribution given by $Q_{fiss.}=BE(A_1,Z_1)+BE(A_2,Z_2)
-BE(A,Z)$ for $^{236,250,260}$U and $^{232,250,256}$Th as a function of $A_1$
fragment in the binary decay $A\rightarrow A_1+A_2$.
The binding energy used for calculation of $Q-$value is
taken from  \cite{moller95} and Refs. \cite{nayak99}
for FRDM and INM, respectively. The fission yield decreases
drastically with increase or
decrease of mass number of given element. Therefore, we have shown
the distribution in the range 90 to 100$\%$ of the peak value in each case.
The vertical line marks the
neutron drip-line for the corresponding element in each panel.
The full and dashed line denote the calculations with FRDM and INM
mass models respectively.
}
\end{figure}

\section{Summary and Conclusions}

We have attempted to understand the mechanism of fission decay by
following the static fission path, and concentrating on the ultimate
shape of the nucleus it likely to acquire, before the break up, i.e., the
neck configuration. This picture  envisaged since the very 
early days of nuclear physics viewing the nucleus to be a classical liquid, is reafirmed here
in a microscopic study carried out in the framework of established nuclear
theory employing welknown many-body nuclear Hamiltonian. The
RMF theory which has gained the confidence of nuclear
communities as a viable theory has been adopted in the present study.
In our calculation, it has been possible to access such highly deformed 
configuration as neck by using very large basis consisting of as many as
28 oscillator states, while for ground-state 10 or 12 states are adequate. 
This study has bared the anatomy of neck, revealing rich neutron-proton
ratio, which progressively increases with the neutron number in the isotope. 
We have been able to find out the length of the neck and its neutron-proton
composition in term of their numbers. The maximum neutron to proton ratio
found by us is 5 which may correlate with the quasi-bound/resonance state
of the heaviest hydrogen isotope $^6$H known so far. This neck with higher
neutron to proton ratio then 6 are not likely to be formed. This suggests that
heavier isotopes of Uranium than $^{260}$U may undergo fission without
formation of neck or more likely fission is inhibited. This may be a 
new feature of fission to be found
in ultra-neutron-rich Uranium isotopes which signals the breakdown of
liquid-drop picture. 

In our investigation, besides the
welknown actinide nuclei $^{236}$U and $^{232}$Th in the valley of 
stability, their highly 
neutron-rich isotopes $^{250}$U, $^{256}$U, $^{260}$U, $^{240}$Th,
$^{250}$Th, $^{256}$Th have been included with the objective of 
finding the fission properties of such exotic nuclei of relevance
in stellar evolution. Some of these nuclei with lower mass number
are likely to be synthesised
in the RIB facilities. The highly neutron-rich necks found in the
calculation in the above exotic nuclei, points out a new mode of fission
decay where along with the two heavy fragments some neutrons will be emitted
simultaneously at scission.
These are extra neutrons in addition to the usual
neutron multiplicity emitted by the two fragments later in the
flight. Thus total multiplicity gets more than doubled of the
expected usual one.  
Due to extreme neutron-richness of the neck it can not be
sucked into the two fragments at scission, but itself breaks down
contributing these neutrons. This inference is supported by
our fission mass-yield study carried out using Q-value systematics.
It will have serious implication in the
energy generation process in the r-process nucleosynthesis in stellar
evolution. Whether this process will be of any utility in the laboratory
is too premature to speculate at the moment.

\section{Acknowledgments}

This work is supported in part by Council of Scientific $\&$ Industrial 
Research (No.03 (1060) 06/EMR-II), as well as the
Department of Science and Technology, Govt. of India, project No. SR/S2/HEP-16/2005.

\end{document}